\documentclass[showpacs]{revtex4}
\emergencystretch=\maxdimen
\usepackage{graphicx}
\usepackage{dcolumn}
\usepackage{amsmath}
\usepackage[latin1]{inputenc}
\usepackage{graphicx, psfrag}
\usepackage{amssymb}
\usepackage[colorlinks=true, citecolor=blue, urlcolor = blue, linkcolor= red, bookmarks=true]{hyperref}
\usepackage{float}
\usepackage{amsmath}
\usepackage{amsfonts}
\usepackage{dcolumn}
\usepackage{hyperref}
\usepackage{subfigure}
\usepackage{pgfplots}
\usepackage{epstopdf}
\usepackage{booktabs}

\makeatletter
\def\btt#1{\texttt{\@backslashchar#1}}
\DeclareRobustCommand\bblash{\btt{\@backslashchar}} \makeatother

\makeatletter
\def\btt#1{\texttt{\@backslashchar#1}}
\DeclareRobustCommand\bblash{\btt{\@backslashchar}} \makeatother

\begin{document}
\title{Shadow and deflection angle of rotating black hole in asymptotically safe gravity  }
\author{Rahul Kumar$^{a}$}\email{rahul.phy3@gmail.com}
\author{Balendra Pratap Singh$^{a}$}\email{balendra29@gmail.com}

\author{Sushant~G.~Ghosh$^{a,\;b}$} \email{sghosh2@jmi.ac.in, sgghosh@gmail.com}
\affiliation{$^{a}$ Centre for Theoretical Physics, Jamia Millia
Islamia, New Delhi 110025, India}
\affiliation{$^{b}$ Astrophysics and Cosmology
Research Unit, School of Mathematics, Statistics and Computer Science, University of
KwaZulu-Natal, Private Bag 54001, Durban 4000, South Africa}

\begin{abstract}
We analytically investigate the shadows cast by rotating black holes in the  asymptotically safe gravity (ASG) by deriving complete null geodesics and observables using the Hamilton-Jacobi equation and Carter separable method. It turns out that the apparent shape and size of the shadow depend on the ASG parameters ($\zeta, \gamma$) in addition to other black hole parameters ($M, a$). The size of black hole shadows monotonically decrease and shadows get more distorted with increasing values of ASG parameters, when compared with the Kerr black hole shadows. In turn, we use shadow observables to estimate black hole spin and ASG parameters. Noteworthy, we find that the deflection angle of the light has been modified by ASG parameters to generalize the Kerr deflection angle, and the corrections in deflection angle are of $\mathcal{O}(\mu$as). In the vanishing limits of ASG parameters our results smoothly reduced to the Kerr black holes. The inferred circularity deviation $\Delta C\leq 0.10$ for the M87* black hole shadow merely constrains the ASG parameter $\zeta$, however, shadow angular diameter $\theta_d=42\pm 3\, \mu$as, within the $1\sigma$ region, places bounds $\zeta\leq 0.1324$ for $\gamma=0.10$.
\end{abstract}
\maketitle
\section{Introduction}
Einstein formulated the theory of General Relativity (GR) long back in 1915, which is still considered as one of the most profound physical theory of all times. GR is perturbatively non-renormalizable (except for pure gravity with no interaction with scalar fields at one-loop level) \cite{Birrell:1982ix, Aharony:1998tt,Hooft}, therefore it is merely considered as an effective field theory and leads difficulty to develop a fully consistent theory of quantum gravity. Nevertheless, we can put a suitable cut-off at high energy limit, which as a result gives a correct description of gravity only up to a certain energy scale and length scale, viz. the classical description of GR cease to validate  near singularity of black holes and demands a new physical theory.\\
The consistent theory of quantum gravity is phenomenologically important at least for two crucial reasons, namely, for the unified theory of fundamental interaction, viz., GUT, and for the understanding of origin of the Universe \cite{Georgi:1974sy, Albrecht:1982wi}. However, in the process of formulation of such theories using perturbative calculations the intangible hurdle comes in the form of non-renormalizability. Particularly this inevitable problem is the manifestation of the fact that the gravitational coupling constant is a dimensionful quantity (dimension $[M]^{-2}$), rendering the infinite series of counter-terms to eliminate the divergences in the theory \cite{Goroff:1985th}. Indeed at each loop correction in perturbation theory, the ultraviolet (UV) divergences in quantum gravity get worse and worse, as a result the divergence has unbounded growth. To tackle this problem of non-renormalizability and to construct a full UV complete quantum theory of gravity physicist have explored various possibilities, e.g., string theory \cite{green2012superstring, Dienes:1996du}, loop quantum gravity \cite{ DeWitt:1967ub,Rovelli:1997yv} and noncommutative geometry \cite{Moffat:2000gr}. The UV completion of gravity would not only validate it to all energy scale (including arbitrary high energy scale) but also demands theory to be renormalizable. Even considering the GR as low energy limit of a more general theory called supergravity \cite{VanNieuwenhuizen:1981ae}, the problem of non-renormalizability could not be resolved completely, although it alleviates the UV  divergence of theory by reducing the number of diverging terms due to the underlying supersymmetry. The quantization of gravity was ended up in an impasse until a non-perturbative approach for renormalization of gravitational theory discovered called the \textit{asymptotic safety} \cite{S. Weinberg,Reuter:1996cp}. It was suggested that there exists a fixed point in the UV limit in the renormalization group flow \cite{ Souma:1999at, Lauscher:2001ya}; the running of a gravitational coupling constant approach fixed point in UV limit such that physical quantities become safe of unphysical divergences. The resulting theory would be asymptotically safe in the sense that at high-energies unphysical divergences are likely to be absent \cite{Stelle:1976gc, Benedetti:2009rx}. The asymptotically safe gravity (ASG) extends the outreach of effective field theory approaches, such that we can safely remove the UV cut-off from the theory and can have a complete description at all energy scale (including arbitrarily high energies) \cite{ Reuter:1996cp, Niedermaier:2006ns, Litim:2003vp,Kawai:1993mb}. \\
Since the ASG has significant effects at short distances (or high energy scale) and black holes provide a natural testbed to study the gravity in its strongest regime, therefore, it would also be interesting to study the light propagation and gravitational deflection angle in the ASG background both from the phenomenological view to understand the theory of quantum gravity and from the observational perspective. Moreover, the bending of light around the black hole play a crucial role in casting its shadow. Black hole solutions in the ASG theory have been extensively studied \cite{Cai:2010zh, Koch:2014cqa, Falls:2010he, Yang:2015sfa, Bonanno:2000ep, Haroon:2017opl}, including their thermodynamics stability \cite{Ma:2014zia}.

The possibilities of ASG signature in the astrophysical black hole spacetimes, where the spacetime curvature effects are strong, are still open and will have far-reaching consequences for our understanding of gravity. Here, we wish to understand what impact the quantum corrections from ASG can have on the morphology of light deflection angle, supermassive black hole shadows, and on the near horizon spacetime geometry, in the context of ongoing observations by the Event Horizon Telescope (EHT). Recently, using the Gauss-Bonnet theorem it is shown that the deflection angle of light can be estimated by integrating the Gaussian curvature of the optical metric over a two-dimensional surface of light propagation \cite{Gibbons:2008rj}. This prompt the study of gravitational lensing theory, moreover, considering source and observer at finite distances from black hole, the deflection angle of light in the weak-field limit has been estimated for varieties of black hole spacetimes \cite{Ishihara:2016vdc,Ishihara:2016sfv, Ono:2017pie, Haroon:2018ryd, Ovgun:2018tua, Jusufi:2018kry, Jusufi:2018jof,Kumar:2019pjp,Kumar:2020pol}. In addition, from the perspective of black hole shadow observations, two major projects, namely, EHT \cite{Doeleman:2017nxk} and BlackHoleCam \cite{Goddi:2017pfy} are targeting supermassive black holes Sagittarius A* (Sgr A*) and M87*. The first observational outcome from the EHT collaboration has already revealed the horizon-scale image of M87* black hole \cite{Akiyama:2019cqa,Akiyama:2019brx,Akiyama:2019sww,Akiyama:2019bqs,Akiyama:2019fyp,Akiyama:2019eap}. Therefore, a rational study of black hole shadow may open a window to probe the quantum gravity. In a pioneering work, Synge \cite{Synge:1966} and Luminet \cite{Luminet:1979}, for the first time ever determined the shadow of Schwarzschild black hole over a bright background, which for Kerr black hole was later determined by Bardeen \cite{Bardeen}. Indeed, shadow is a gravitationally lensed image of the photon captured region accounting for those photons which inevitably end up in the black hole over a finite affine parameter, thus, its silhouette appears as a sharp boundary between dark and bright regions. Providing the black hole shadow relevance for testing the nature of gravity in the strong-field regime, a large comprehensive literature has addressed the shadow study either within GR or modified gravities \cite{De,Amarilla:2010zq,Amarilla:2013sj,Yumoto:2012kz,Abdujabbarov:2016hnw,Amir:2016cen,Tsukamoto:2014tja,Bambi:2010hf,Goddi:2016jrs,Takahashi:2005hy,Wei:2013kza,Abdujabbarov:2012bn,Amarilla:2011fx,Bambi:2008jg,Atamurotov:2013sca,Wang:2017hjl,Kumar:2017vuh,Schee:2008kz,Grenzebach:2014fha,Liu:2020ola,Jusufi:2019nrn,Jusufi:2020cpn,Kumar:2020yem}. In addition, the effect of surrounding thin plasma medium on the rotating black hole shadow has also been studied \cite{Perlick:2015vta,Atamurotov:2015nra,Abdujabbarov:2015pqp}, the Kerr black hole shadow with scalar hair by ray tracing method was reported in \cite{Cunha:2015yba,Cunha:2016bpi} and it is also extended  to the higher dimensional spcaetime \cite{Papnoi:2014aaa,Singh:2017vfr,Amir:2017slq,Abdujabbarov:2015rqa}. The characterizations of shadow also provide a potential way to extract the black hole parameters \cite{Hioki:2009na, Abdujabbarov:2015xqa,Tsupko:2017rdo, Kumar:2018ple}. Recently, it is found that shadow observations may offer a way to distinguish noncommutative geometry inspired black hole from the Kerr black hole \cite{Wei:2015dua}. It is also believed that quantum modifications to black hole solutions may propagate to the horizon scale and may have observable implications for astrophysical phenomena that originate in the vicinity of horizons \cite{Maldacena:2013xja, Giddings:2011ks}. Indeed, metric fluctuations have similar observable effects on  black hole shadows \cite{Giddings:2016btb}.  Moreover, a detailed comprehensive analysis of black hole shadow can provide a better opportunity to study the physical processes at the very vicinity of horizon and also to probe any signatures of quantum effects on gravity in its strongest regime. It is the purpose of this paper to investigate the effects of quantum corrections, owing due to the ASG theory, on the gravitational lensing and shadows of rotating ASG black hole spacetimes.  \\
The paper is organized as follows. In Section~\ref{assect2}, we discuss the null geodesics in the rotating ASG black hole spacetime. Further in Section~\ref{assect3}, we obtain the black hole shadow and studied the gravitational deflection of light in Section~\ref{assect4}.  In Section~\ref{assect5}, we summarize our main results.
 
\section{Null geodesics around rotating ASG black holes}\label{assect2} 
In the ASG theory, the Newton's constant $G$ is taken as $r$ dependent function $G(r)$, which evolves under
the renormalization group equations for gravity. The static, spherically symmetric black hole solution, using the renormalization flow equation, in ASG was reported in Ref. \cite{Bonanno:2000ep}. Later, using the running gravitational coupling constant as \cite{Bonanno:2000ep,Falls:2010he}
\begin{equation}\label{Gr}
G(r)=\frac{G_0 r^3}{r^3+\zeta(r+\gamma MG_0 )},
\end{equation}
Torres \cite{Torres:2017gix} obtained the rotating black hole metric by using the Newman-Janis algorithm \cite{Newman:1965tw}. Though this algorithm was developed within general relativity, it has been more recently applied to non-rotating solutions in modified gravity theories, e.g., some references \cite{Johannsen:2011dh,Jusufi:2019caq,Bambi:2013ufa,Ghosh:2014hea,Moffat:2014aja,Hansen:2013owa}. The quantum improved rotating black hole metric in ASG theory, in Boyer-Lindquist coordinates, takes the following form \cite{Torres:2017gix}
 \begin{eqnarray}
 ds^2&=& - \left(1-\frac{2MG(r)r}{\Sigma}\right)dt^2-\frac{4MG(r)ar\sin^2\theta}{\Sigma}dtd\phi+\frac{\Sigma}{\Delta}dr^2+\Sigma d\theta^2\nonumber\\
  &&+\frac{\left((r^2+a^2)^2-a^2\Delta\sin^2\theta\right)\sin^2\theta}{\Sigma}{d\phi}^2,\label{RotMet}
  \end{eqnarray}
  with
  \begin{equation}
  \Sigma=r^2+a^2\cos^{2}{\theta}, \quad \Delta=r^{2}+a^2-2MG(r)r.
  \end{equation}
Here, $\zeta$ and $\gamma$ are new variables of ASG theory and $G(r)$ is a running (or scale dependent) coupling constant, which in the infra-red (IR) limit, $r\gg \sqrt{\zeta}$, reduces to
\begin{equation}
G(r)\simeq G_0\left(1-\frac{\zeta}{r^2}\right),\label{Gr2}
\end{equation}
and thereby the metric (\ref{RotMet}) with (\ref{Gr2}) simplifies to rotating ASG black hole metric in the IR limit \cite{Haroon:2017opl,Reuter:2010xb}.
Since, $G(r)$ in Eq.~(\ref{Gr}) at asymptotically large $r$ matches with the Newton's gravitational constant $G_0$, i.e., $r\to \infty\Rightarrow G(r)\rightarrow G_0$, the metric (\ref{RotMet}) reduces to that of Kerr black hole. Before studying the black hole shadow in ASG, it is necessary to discuss the test particle trajectory in the rotating black hole spacetimes. Lets consider a black hole in luminous background, the light coming from the source will get deflected due to the strong gravitational field near black hole before reaching to the distant observer. Depending upon the deflection, photons trajectories can be classified into three categories, namely, capture orbit, scattering orbit, and unstable circular orbit \cite{Hioki:2009na,Amarilla:2013sj}. Photons moving on unstable circular orbits account for the apparent photon ring around the black hole. The Hamilton-Jacobi equation, in terms of the action $\mathcal{S}=\mathcal{S}(x^{\alpha},\tau)$ as a function of spatial coordinates $x^{\alpha}$ and affine parameter along geodesics $\tau$, reads  \cite{Carter:1968rr}
\begin{eqnarray}
\label{HmaJam}
\frac{\partial {\mathcal{S}}}{\partial \tau} = -\frac{1}{2}g^{\alpha\beta}\frac{\partial {\mathcal{S}}}{\partial x^\alpha}\frac{\partial {\mathcal{S}}}{\partial x^\beta}.
\end{eqnarray}
The stationary and axially symmetric spacetime described in Eq.~(\ref{RotMet}) admits the time-translational and rotational invariance, which guarantees the existence of two conserved quantities associated with the test particle along geodesics namely, energy $\mathcal{E}$ and axial angular momentum $\mathcal{L}$. Therefore, we can choose a separable solution in the following form \cite{Chandrasekhar:1992}
\begin{eqnarray}\label{www}
\mathcal{S}=\frac12 {m_0}^2 \tau -{\cal E} t +{\cal L} \phi +\mathcal{S}_r(r)+\mathcal{S}_\theta(\theta) \label{action},
\end{eqnarray}
where $m_0$ is the rest mass of the test particle which is zero for photons. Substituting Eq.~(\ref{www}) in Eq.~(\ref{HmaJam}) and applying the variable separable method, we analytically derived the null geodesic equations around rotating ASG black hole as 
\begin{eqnarray}
\Sigma \frac{dt}{d\tau}&=&\frac{1}{\Delta}\left[\left((r^2+a^2)^2-a^2\Delta\sin^2\theta\right){\cal E}-2MG(r)ar{\cal L}\right]  ,\label{tuch}\\
\Sigma \frac{dr}{d\tau}&=&\sqrt{{\mathcal{R}(r)}}\ , \quad {\mathcal{R}(r)}\geq 0  \label{r}\\
\Sigma \frac{d\theta}{d\tau}&=&\sqrt{\Theta(\theta)}\ ,\quad \Theta(\theta)\geq 0 \label{th}\\
\Sigma \frac{d\phi}{d\tau}&=&\frac{1}{\Delta}\left[2MG(r)ar{\cal E}+(\Sigma-2MG(r)r){\cal L} \csc^2\theta\right]\ ,\label{phiuch}
\end{eqnarray}
where $\mathcal{R} (r)$ and ${\Theta}(\theta)$ in Eq.~(\ref{r}) and (\ref{th}) are, respectively, the sole functions of radial and angular coordinate, and have the following forms
\begin{eqnarray}\label{06}
\mathcal{R}(r)&=&\left[(r^2+a^2){\cal E}-a{\cal L}\right]^2-(r^{2}-2MG(r)r+a^{2})[(a{\cal E}-{\cal L})^2+{\cal K}],\quad \\ 
\Theta(\theta)&=&{\cal K}-\left({\cal L}^2\csc^2\theta-a^2{\cal E}^2\right)\cos^2\theta,
\end{eqnarray}
with the separable Carter constant $\mathcal{K}$. Equation.~(\ref{tuch})-(\ref{phiuch}) completely define the photon geodesics around the rotating ASG black holes. Next, we define two dimensionless impact parameters in terms of constants of motion as $\xi=\mathcal{L}/\mathcal{E}$ and $\eta=\mathcal{K}/\mathcal{E}^2$. From the radial equation of motion  (\ref{r}) we can rewrite the function $\mathcal{R}(r)$ for photon case in terms of these impact parameters as
\begin{equation}
\mathcal{R}(r)=\frac{1}{{\cal{E}}^2}\left[\left((r^2+a^2) -a{\xi}\right)^2-(r^{2}-2MG(r)r+a^{2})\left((a -{\xi})^2+{\eta}\right)\right].
\end{equation} 
Noticeably, $\cal{R}$$(r)$ and ${\Theta}(\theta)$ are related to the effective potential, respectively, in $r$ and $\theta$ directions, as
\begin{equation}
\frac{1}{2}\left(\frac{dr}{d\tau}\right)^2=\mathcal{E}^2-V_{eff}(r),\qquad \frac{1}{2}\left(\frac{d\theta}{d\tau}\right)^2=\mathcal{E}^2-V_{eff}(\theta).\label{pot}
\end{equation} 
From Eqs.~(\ref{r}), (\ref{th}) and (\ref{pot}), we get the functional form of effective potentials experienced by photons in a black hole spacetime. Depending on the values of impact parameters, photons can move in the bound orbits, which are characterized by the maximum of radial effective potential. The photons moving with higher angular momentum than those moving on unstable orbits feels scattering, while those with smaller angular momentum fall into the black hole \cite{Chandrasekhar:1992}. Therefore, the unstable orbits, which separate the scattering and capture orbits are crucial for the black hole shadow. Thus photons which account for the black hole shadow  experience radial turning point in their trajectory and also correspond for local maximum of radial effective potential: 
\begin{equation}
 \mathcal{R}(r)=\frac{\partial \mathcal{R}(r)}{\partial r}=0,\quad \text{and}\quad \frac{\partial^2 \mathcal{R}(r)}{\partial r^2}\geq 0.\label{asvr} 
\end{equation} 
We solve Eq.~(\ref{asvr}) to obtain the constant of motion $\eta$ and $\xi$, characterizing these unstable orbits 
\begin{eqnarray}
\eta&=&\frac{-{r_0}^4}{P^2}\big[4 a^2 M ({r_0}^3 + {r_0} \zeta 
       + M \gamma\zeta)^2(-{r_0}^3 + {r_0} \zeta+ 2 M \gamma \zeta) +({r_0}^5 (-3 M + {r_0}) + 
  {r_0}^3 (2 {r_0} \nonumber\\
  &&+ M (-1 + 2 \gamma)) \zeta +({r_0} +M \gamma)^2 \zeta^2)^2  \big], \label{asxiexp}\\
\xi &=& \frac{1}{P}\big[{r_0}^5 ({r_0}^2 (-3 M{r_0})+ a^2 (M + {r_0}))+{r_0}^2 (a^2 ({r_0} (3 M + 2 {r_0}) + 2 M (2 M + {r_0}) \gamma) + 
    {r_0}^3 (2 {r_0}\nonumber\\
    && + M (-1 + 2 \gamma))) \zeta+ (a^2 + {r_0}^2) ({r_0} + 
    M \gamma)^2 \zeta^2\big],\label{asetaexp}
\end{eqnarray} 
where $P=a \left((M - {r_0}) {r_0}^5 + 
{r_0}^2 (-2 {r_0}^2 + M {r_0} (3 - 2 \gamma) + 4 M^2 \gamma) \zeta - ({r_0} +
M \gamma)^2 \zeta^2\right)$ and $r_0$ is the unstable orbit radius. The trajectory of bound photon orbits around the black hole in ASG is defined by the above Eqs.~(\ref{asxiexp}) and (\ref{asetaexp}), which in the limit,  $(\zeta \to 0)$  reduces  to
\begin{eqnarray}
\xi &=& \frac{{r_0}^2 ( {r_0}-3 M) + a^2 (M + {r_0})}{a (M - {r_0})}, \label{xiexp}\\
\eta &=&\frac{{r_0}^3 (4 a^2 M -{r_0} ({r_0}-3 M)^2)}{a^2 (M - {r_0})^2},
\end{eqnarray}

\begin{figure}
\begin{tabular}{c c c c}     
	    \includegraphics[scale=0.8]{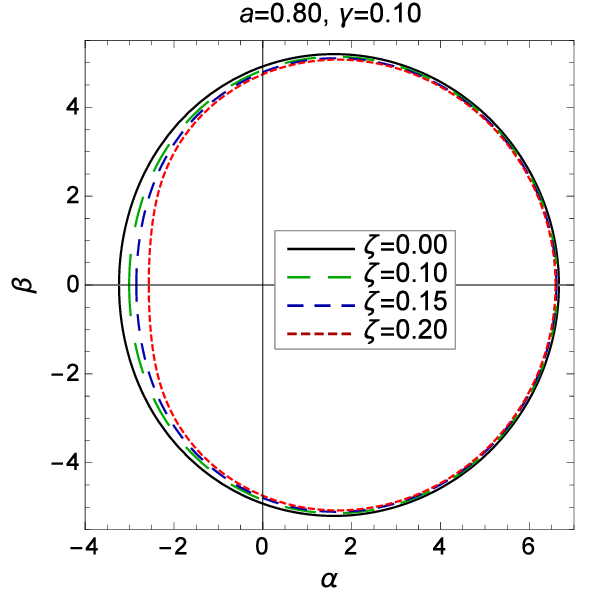}&
		\includegraphics[scale=0.8]{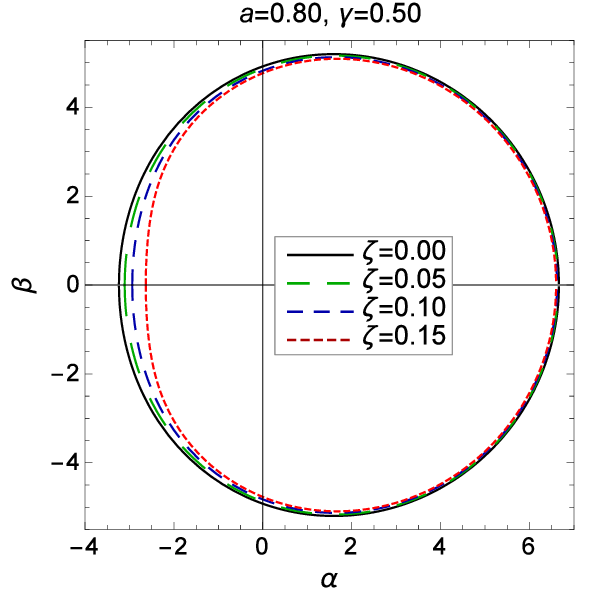}\\
		\includegraphics[scale=0.8]{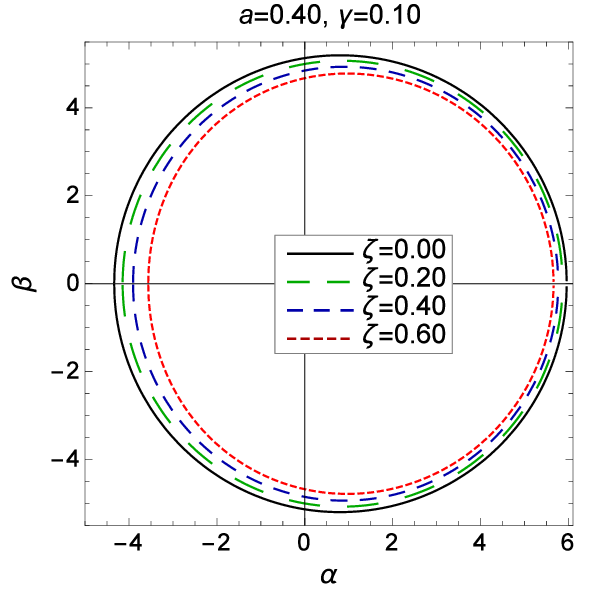}&
		\includegraphics[scale=0.8]{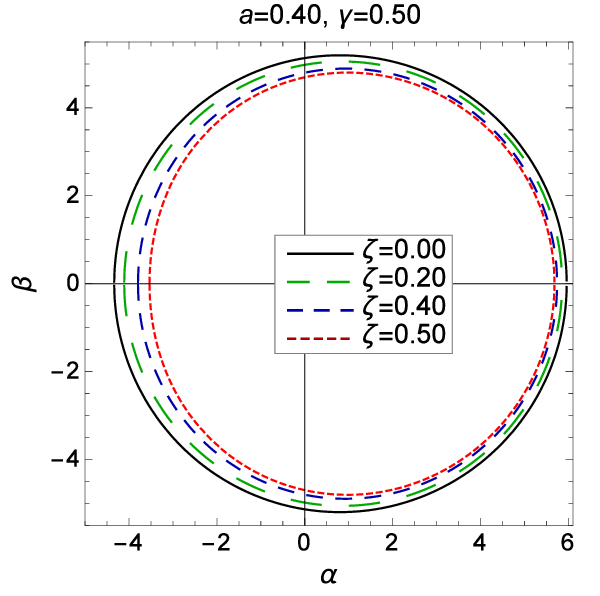}
\end{tabular}
	  \caption{Rotating ASG black hole shadows for different values of $\gamma$ and $a$ and varying  $\zeta$, black solid curves correspond to the Kerr black hole shadow.}\label{asfig2}
\end{figure}
and corresponds for the Kerr black hole \cite{Hioki:2009na}. The expressions $\eta$ and $\xi$ in Eqs.~(\ref{asxiexp}) and (\ref{asetaexp}) give the critical values of impact parameters for photon undergoing unstable orbits around rotating ASG black hole. 
\section{Black hole shadow \label{assect3}}
 Black hole shadow is better visualized by celestial coordinates $\alpha$ and $\beta$, defined as \cite{Hioki:2009na} 
\begin{equation}\label{aalp}
 \alpha=\lim_{r_*\rightarrow\infty}\left(-r_*^2 \sin{\theta_0}\frac{d\phi}{d{r}}\right),\quad  \beta=\lim_{r_*\rightarrow\infty}r_*^2\frac{d\theta}{dr},
\end{equation}
where $r_*$ is the distance between the observer and the black hole while $\theta_0$ is the inclination angle between the line of sight of the observer and the rotational axis of the black hole. Here, we assume observer at infinity. Using geodesic equations (\ref{tuch})-(\ref{phiuch}) and celestial coordinates (\ref{aalp}), one can directly relate coordinates $\alpha$ and $\beta$ to the impact parameters by   
 \begin{equation}
\alpha= -\frac{\xi}{\sin\theta_0},\qquad  \beta=\sqrt{\eta+a^2\cos^2\theta_0-\xi^2\cot^2\theta_0 }.
\end{equation}
The effect of spin parameter on black hole shadow is most significant for inclination angle $\theta_0 =\pi/2$ and decreases further as we move towards a polar end. Considering an observer at the equatorial plane, celestial coordinates $\alpha$ and $\beta$ reduce to
\begin{equation}
\alpha=-\xi,\qquad \beta=\pm\sqrt{\eta},\label{pt}
\end{equation} 
which satisfy the following relation
\begin{eqnarray}
\alpha^2+\beta^2&=& \frac{1}{P^2}\Big[\Big(( 2 \gamma  \zeta  M-{r_0}^3+\zeta  {r_0})({r_0}^4\left(4 a^2 M \left(\gamma  \zeta  M+{r_0}^3+\zeta  {r_0}\right)^2\right)\Big)+\Big(\Big({r_0}^5 ({r_0}-3 M)\nonumber\\
&&+\zeta  {r_0}^3(2 \gamma -1) M+ 2 \zeta {r_0}^4)+\zeta ^2 (\gamma  M+{r_0})^2\Big)^2\Big)+\Big({r_0}^5 ({r_0}^2 (-3 M + {r_0})\nonumber\\
&& + a^2 (M + {r_0})) +{r_0}^2\Big(a^2\zeta ^2 \left(a^2+{r_0}^2\right) (2 \gamma  M (2 M+{r_0})+{r_0} (3 M+2 {r_0}))\Big)(\gamma  M+{r_0})^2 \nonumber\\
&&+ {r_0}^3 ((2 \gamma -1) M+2 {r_0}))\zeta\Big)^2\Big].
\end{eqnarray}
\begin{figure*}
\begin{center}
    \includegraphics[scale=0.8]{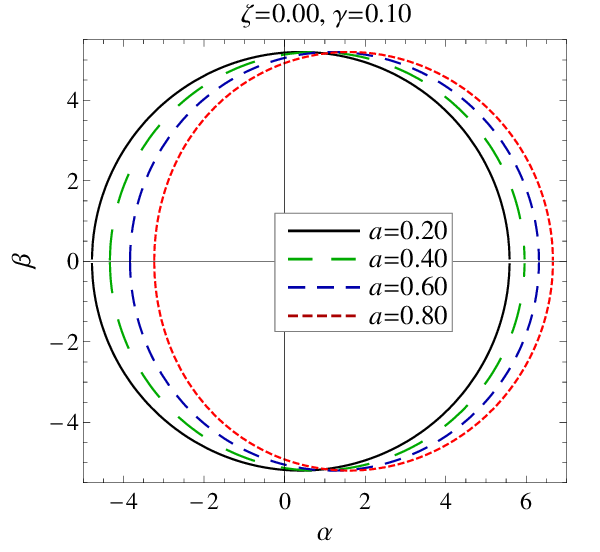} 
	\includegraphics[scale=0.8]{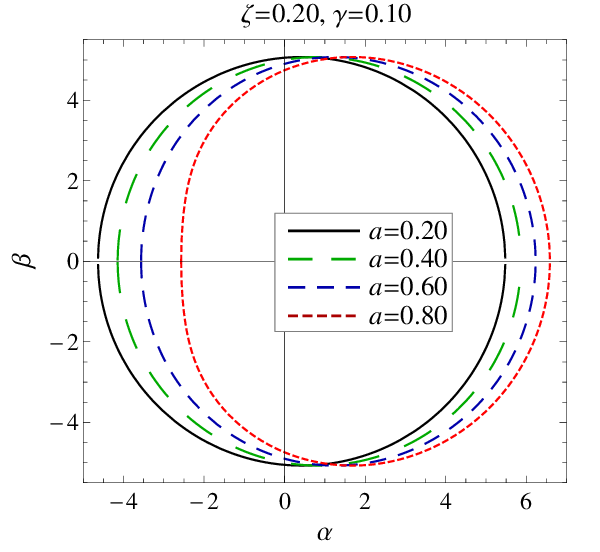} \\
	\includegraphics[scale=0.8]{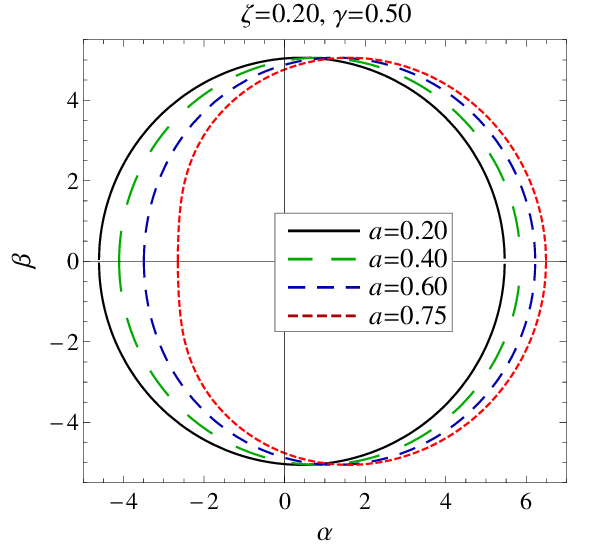} 
	\includegraphics[scale=0.8]{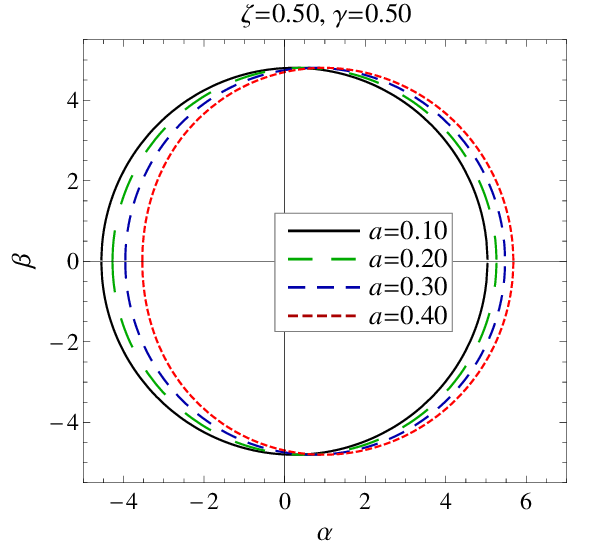} \\
	\includegraphics[scale=0.8]{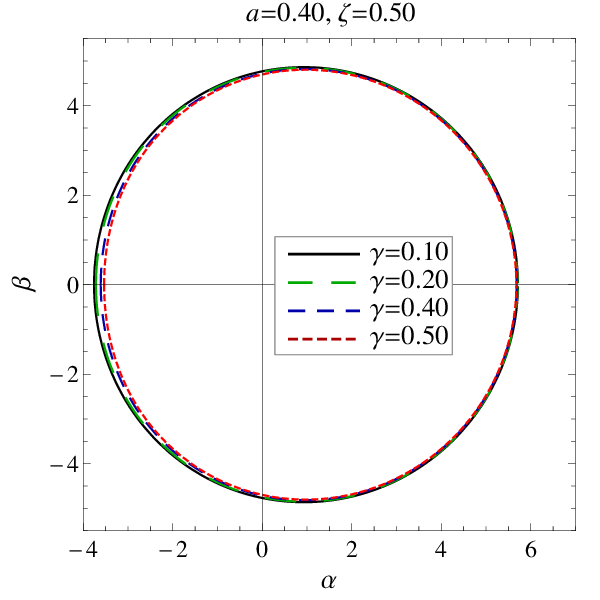} 
	\includegraphics[scale=0.8]{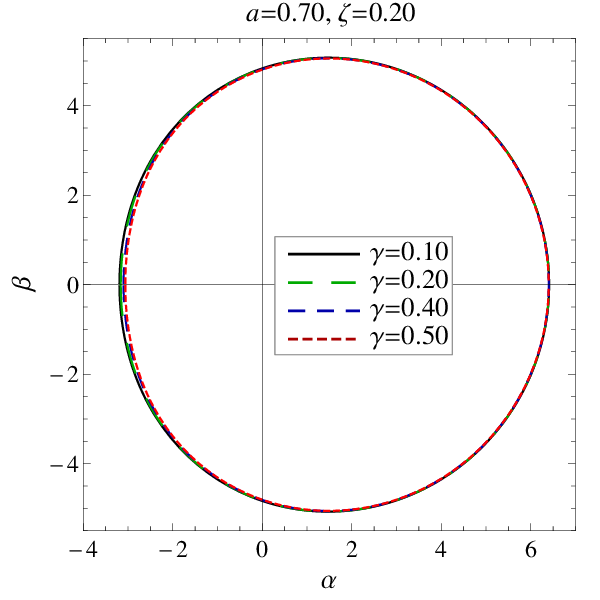} 
	 \end{center}   
    \caption{Rotating ASG black hole shadows for varying $a$ and $\gamma$.}\label{asfig3} 
\end{figure*}
 \begin{figure*}
	\begin{tabular}{c c c c}
		\includegraphics[scale=0.8]{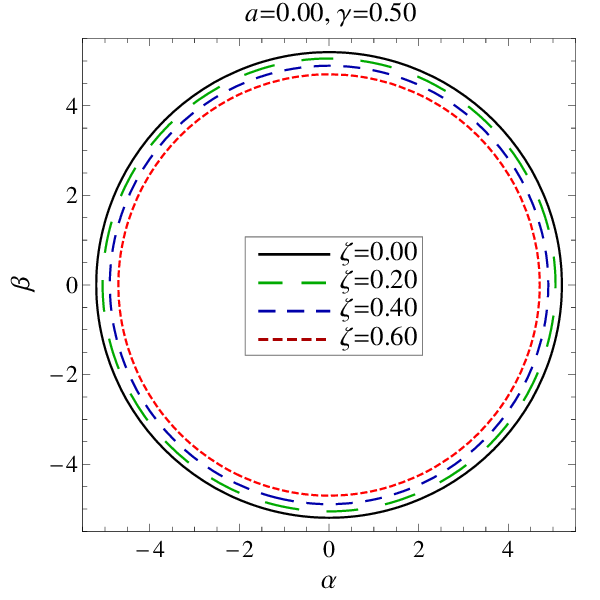} 
		\includegraphics[scale=0.8]{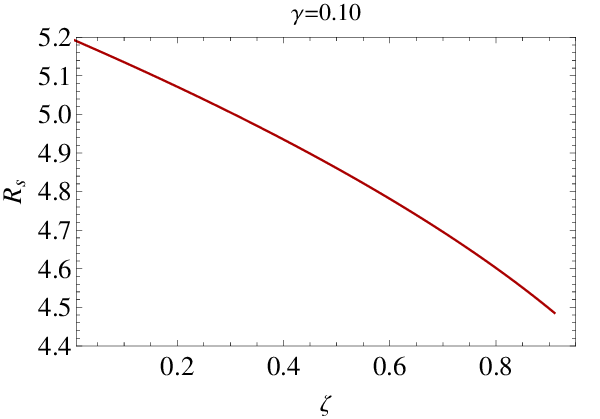} 
	\end{tabular}
	\caption{\label{asfig6} Plots showing the non-rotating ASG black hole shadows and its observable $R_s$ with varying $\zeta$, black solid curve corresponds for the Schwarzschild black hole shadow.}
\end{figure*}
Again in the limit $\zeta\rightarrow0$, we obtain the corresponding equation for the Kerr black hole \cite{Hioki:2009na}
\begin{eqnarray}
{\alpha}^2+{\beta}^2 &=& \frac{a^2 (M+{r_0})^2-6 M^2 {r_0}^2+2 {r_0}^4}{(M-{r_0})^2}.
\end{eqnarray}
We plot $\alpha$ vs $\beta$, in Fig.~{\ref{asfig2}} and Fig.~{\ref{asfig3}}, to show rotating ASG black hole shadows for various values of parameters. In Fig.~{\ref{asfig2}} we plotted rotating ASG black hole shadows for various values of $\zeta$ by fixing $a$ and $\gamma$, while plots in Fig.~{\ref{asfig3}} shows shadows for varying $a$ and $\gamma$ by keeping $\zeta$ fixed. We find that the spin $a$ and ASG parameter $\zeta$ significantly affect the shape and the size of the shadow.
Rotating ASG black hole shadows are considerably different from the Kerr black hole shadows (cf. Fig. \ref{asfig2}). To characterize the shape and size of the shadow, we adopt two observables: shadow radius $R_s$ and distortion parameter $\delta_s$ \cite{Hioki:2009na}. Indeed, $R_s$ is the radius of reference circle passing by three points (top $(\alpha_t,\beta_t)$, bottom $(\alpha_b,\beta_b)$, and extreme right $(\alpha_r,0)$) of the shadow boundary, whereas $\delta_s$ characterizes the shadow deviation from the perfect circle. The first observable $R_s$ can be written in terms of the celestial coordinates of some specific points on the shadow as  follow \cite{Hioki:2009na}
\begin{equation}
R_s=\frac{(\alpha_t-\alpha_r)^2+{\beta_t}^2}{2|{\alpha_t}-{\alpha_r}|}.
\end{equation}
The second observable $\delta_s$ depends upon the dent $D_s (=\alpha_l-\tilde{\alpha}_l)$ and shadow radius $R_s$ which is expressed in the following form \cite{Hioki:2009na}
\begin{equation}
\delta_s=\frac{D_s}{R_s}.
\end{equation} 
Here, $\alpha_l$ and $\tilde{\alpha}_l$ are the abscissa coordinates, where shadow and the reference circle cut the negative $\alpha-$axis.
\begin{figure}
	\begin{tabular}{c c}
		\includegraphics[scale=0.8]{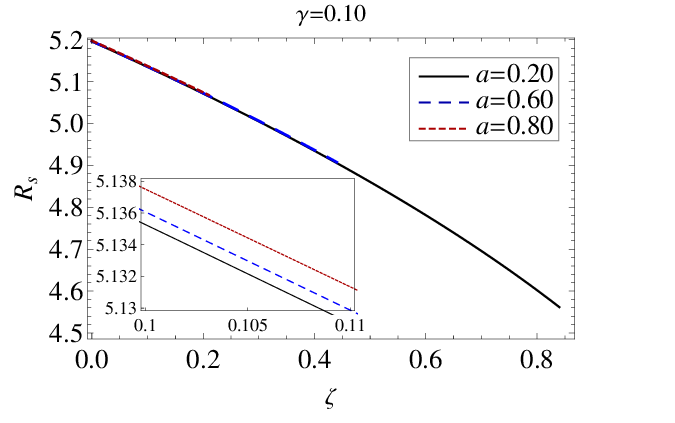} 
		\includegraphics[scale=0.8]{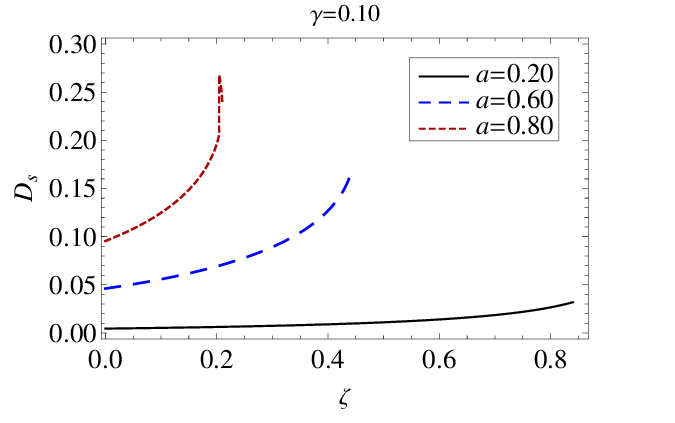}
	\end{tabular} 
	\caption{\label{asfig4} Plots showing the  variation of shadow observables $R_s$ and $\delta_s$ for rotating ASG black hole shadows with varying $\zeta$ and $a$ for $\gamma=0.10$.}
\end{figure}
\begin{figure}
	\begin{center}
		\begin{tabular}{c c}
		\includegraphics[scale=0.84]{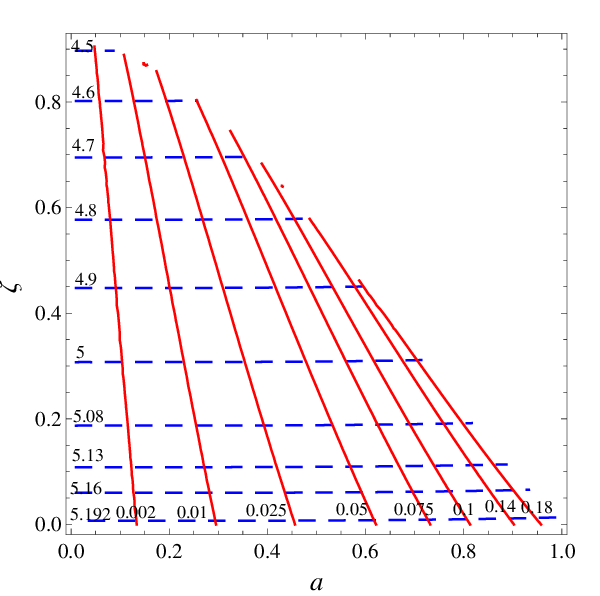} &
		\includegraphics[scale=0.84]{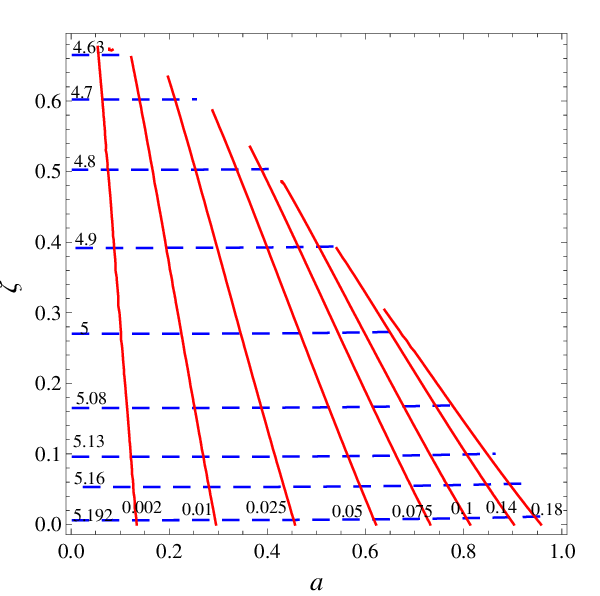}
		\end{tabular}
	\end{center}
	\caption{\label{asfig5} The contours of $R_s$ (dashed blue lines) and $\delta_s$ (solid red lines) in the $(a,\zeta)$ plane for $\gamma=0.10$ (left panel) and  $\gamma=0.50$ (right panel), each curve is labeled with corresponding values of $R_s$ and $\delta_s$. The point of intersection of $R_s$ and $\delta_s$ gives the values of black hole parameters.}
\end{figure}
We have also plotted non-rotating black hole $(a=0)$ shadows in Fig.~\ref{asfig6} for different values of $\zeta$. The first silhouette in Fig.~\ref{asfig6} (full black line) corresponds to the Schwarzschild black hole in GR ($\zeta=0$), and it can be clearly inferred from the figure that with increasing values of $\zeta$ the shadow size decreases. From the study of the non-rotating case, we conclude that the black hole shadow in ASG is still circular but appears smaller as compared to the shadow of the Schwarzschild black holes in standard GR (cf. Fig.~{\ref{asfig6}}). While for the rotating ASG black holes shadow, the apparent size decrease and shadow gets more and more distorted with increasing ASG parameter $\zeta$ (cf.  Fig.~\ref{asfig4}). Figures \ref{asfig2} and \ref{asfig4} infer that the shadows of rotating ASG black holes are smaller and more distorted than those of Kerr black hole. To estimate black hole parameters, we plotted $R_s$ and $D_s$ in the ($a,\zeta$) plane in Fig.~{\ref{asfig5}}, where the intersection of constant curves of $R_s$ and $D_s$ gives the exact values of black hole spin $a$ and ASG parameter $\zeta$.\\

The EHT Collaboration, using the very large baseline interferometry techniques, captured the images of central compact emission region at the center of galaxy M87 at the 1.3 mm wavelength \cite{Akiyama:2019cqa,Akiyama:2019eap,Akiyama:2019bqs,Akiyama:2019fyp}. Though the shadow of M87* black hole  is consistent with that predicted for a Kerr black hole in general relativity, it does not rule out various black hole models in general relativity as well in modified gravities \cite{M87test,Kumar:2019pjp,Kumar:2020pol}. We use shadow observables, namely, the circularity deviation $\Delta C$, and angular diameter $\theta_d$ to constrain the parameter space of rotating ASG black holes. The distortion in the observed shadow of M87* black hole is parameterized by $\Delta C$, such that the estimated deviation is bounded by $\Delta C\leq 10\%$ \cite{Akiyama:2019cqa}. The average shadow radius $\bar{R}$ measured from the geometric center of shadow $(\alpha_G,\beta_G)$ reads as \cite{Johannsen:2010ru}
\begin{equation}
\bar{R}=\frac{1}{2\pi}\int_{0}^{2\pi} R(\varphi) d\varphi,
\end{equation}
where 
\begin{eqnarray}
R(\varphi)&=& \sqrt{(\alpha-\alpha_G)^2+(\beta-\beta_G)^2},\nonumber\\
\alpha_G&=&\frac{\mid \alpha_{max}+\alpha_{min}\mid}{2};\qquad \beta_G=\frac{\mid \beta_{max}+\beta_{min}\mid}{2},\nonumber\\
\varphi&\equiv & \tan^{-1}\left(\frac{\beta}{\alpha-\alpha_G}\right).
\end{eqnarray}
Here, $\alpha_G$ and $\beta_G$ measure the shadow displacement along the $\alpha$ and $\beta-$axis, nevertheless, due to reflection symmetry along the $\alpha-$axes $\beta_G=0$.
The circularity deviation $\Delta C $ is conceived as the root-mean-square distance from the average radius \cite{Johannsen:2010ru,Johannsen:2015qca}
\begin{equation}
\Delta C=2\sqrt{\frac{1}{2\pi}\int_0^{2\pi}\left(R(\varphi)-\bar{R}\right)^2d\varphi}.
\end{equation}
\begin{figure}[h!]
	\begin{tabular}{ c c}
\includegraphics[scale=0.8]{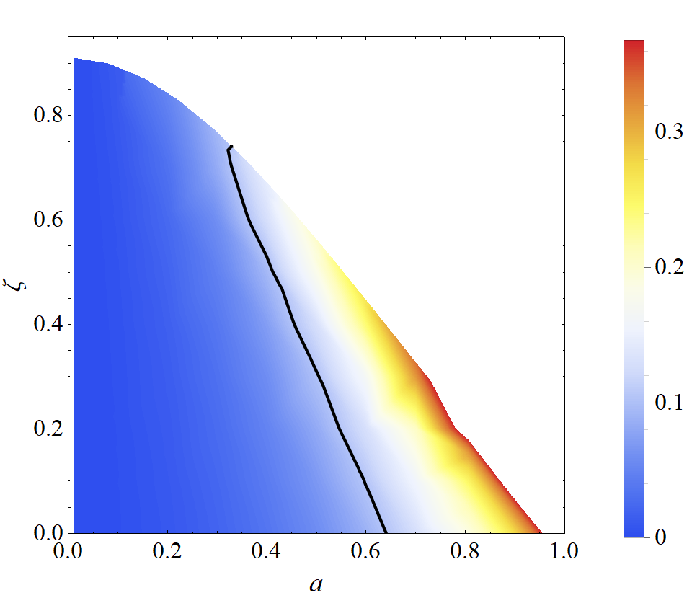}&
\includegraphics[scale=0.8]{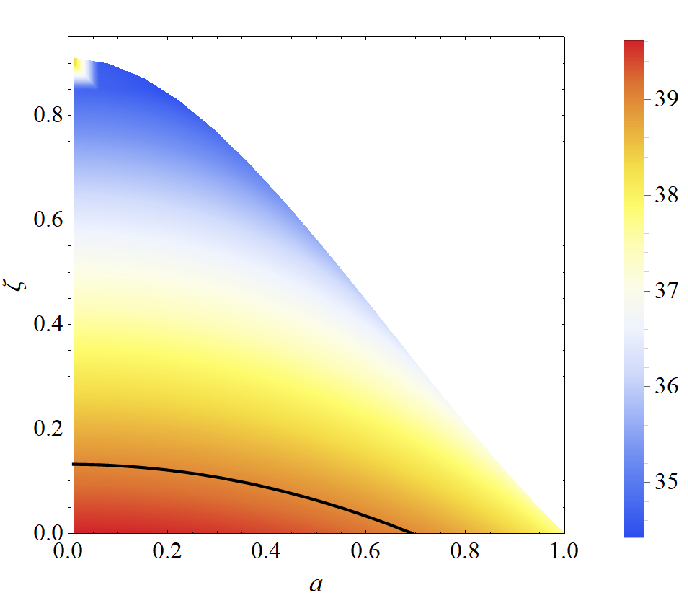}
    \end{tabular}
\caption{Circularity deviation observable $\Delta C$ (left panel) and the angular diameter $\theta_d$ (right panel) as a function of ($a,\zeta$) for the rotating ASG black holes for $\gamma=0.10$. Black solid lines correspond to the M87* black hole shadow bounds, namely $\Delta C=0.10$  and $\theta_d=39\mu$as within the $1\sigma$ region, such that the region above (or to the right-side of) the black line is excluded by the EHT bounds.}\label{M87}
\end{figure}
\begin{figure}
	\begin{tabular}{ c c}
		\includegraphics[scale=0.8]{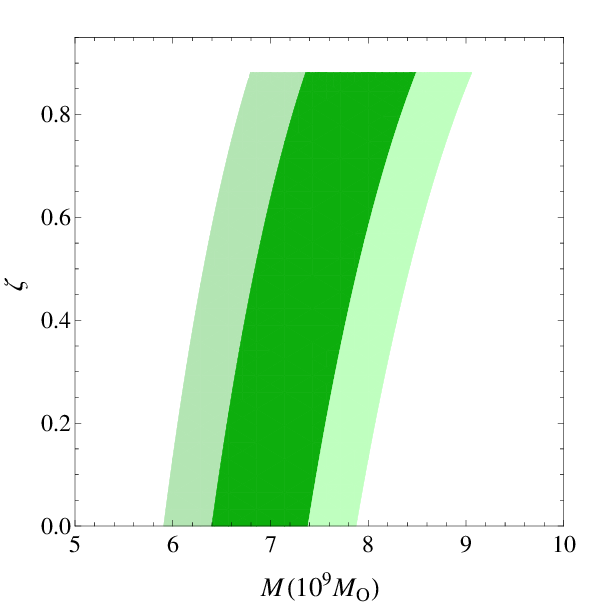}&
		\includegraphics[scale=0.8]{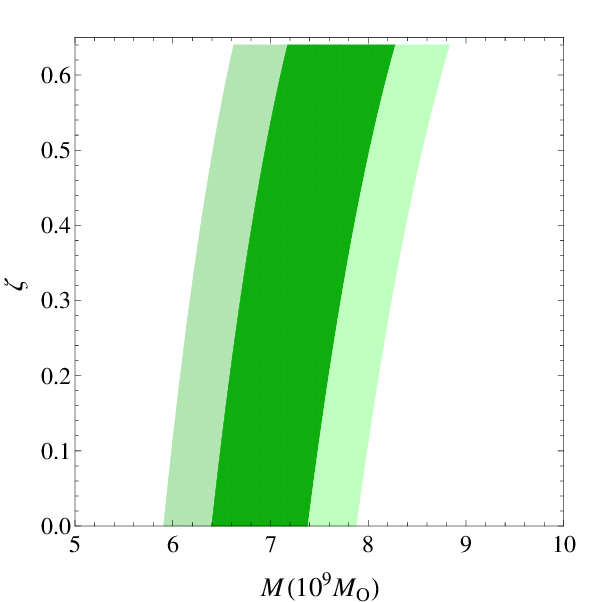}
	\end{tabular}
	\caption{Constraints on the ASG parameter $\zeta$ and estimated black hole mass using M87* shadow angular size within $1\sigma$ (dark green region) and $2\sigma$ (light green region) for $\gamma=0.10$ (left panel) and $\gamma=0.60$ (right panel).}\label{Mass}
\end{figure}
Further, a far distant observer, at a distance $d$ from the black hole, measures the angular diameter $\theta_d$ for the black hole shadow 
\begin{equation}
\theta_d=2\frac{R_{sh}}{d},\;\;\;\; R_{sh}=\sqrt{A/\pi},
\end{equation}    
where $A$ is the shadow area \cite{Kumar:2018ple}. For the M87* black hole shadow the inferred angular diameter is $\theta_d=42\pm 3\, \mu$as \cite{Akiyama:2019cqa}. We calculated the circularity deviation $\Delta C$ and angular diameter $\theta_d$ for the rotating ASG black hole shadows for $M=6.5 \times 10^9 M_{\odot}$, and $d=16.8\, M$pc, and plotted in Fig.~\ref{M87} as a function of $a$ and $\zeta$.  This is evident that for limited parameter space, i.e., the regions enclosed by the black solid line, the rotating ASG black hole shadows are consistent with the M87* shadow (cf.  Fig.~\ref{M87}). However, in contrast to $\Delta C$, the shadow angular diameter $\theta_d$ within the $1\sigma$ region leads to stronger constraint on $\zeta$, viz., $\zeta\leq 0.1324$ for $\gamma=0.10$. Furthermore, using the shadow angular diameter within the $1\sigma$ and $2\sigma$ regions, the constraints on the parameter $M$ and $\zeta$ for $a=0$ and $\gamma=0.10, 0.60$ are shown in Fig.~\ref{Mass}. For $\gamma=0.10$, $1\sigma$ bound on mass is $6.4\times 10^9M_{\odot}\leq M\leq 8.49\times 10^9M_{\odot}$ and $2\sigma$ bound is $5.91\times 10^9M_{\odot}\leq M\leq 9.06\times 10^9M_{\odot}$, whereas for $\gamma=0.60$, the $1\sigma$ bound leads $6.4\times 10^9M_{\odot}\leq M\leq 8.28\times 10^9M_{\odot}$ and $2\sigma$ bound is $5.91\times 10^9M_{\odot}\leq M\leq 8.83\times 10^9M_{\odot}$ (cf. Fig.~\ref{Mass}).

\section{Gravitational deflection of light}\label{assect4}
\begin{figure}
	\includegraphics[scale=0.65]{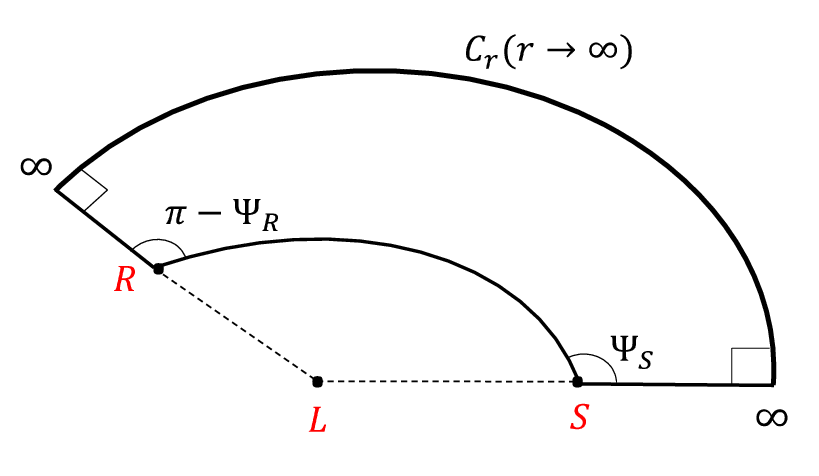}
	\caption{A schematic for lensing setup and domain of integration ${}_R^{\infty}\Box_{S}^{\infty}$.}\label{lensing}
\end{figure}

Next, we will calculate the deflection angle of light in the weak-field limit caused by the rotating ASG black holes by using the Gauss-Bonnet theorem. Using the metric (\ref{RotMet}), we consider that the source ($S$) and observer ($R$) are at finite distance from the lens object (black hole $L$), and define the defection angle \cite{Gibbons:2008rj,Ishihara:2016vdc,Ono:2017pie}:
\begin{equation}
\alpha_D=\Psi_R-\Psi_S+\Phi_{RS},
\end{equation}
where $\Psi_R$  and $\Psi_S$ are the respective angles made by light rays from radial direction at observer and source positions, $\Phi_{RS}$ is the longitude separation angle between observer and source (cf. Fig.~\ref{lensing}). The quadrilateral ${}_R^{\infty}\Box_{S}^{\infty}$ in Fig.~\ref{lensing} is made of light curve from $S$ to $R$, two radial lines joining source and observer to the lens, and a circular arc segment $C_r$ of radius $r_c (r_c\to\infty)$.
Using the null condition $ds^2=0$ for metric (\ref{RotMet}), we solve for $dt$
\begin{equation}
dt= \pm\sqrt{\gamma_{ij}dx^i dx^j}+N_i dx^i,
\end{equation}
where $i,j=1,2,3$. Metric $\gamma_{ij}$ defines a three-dimensional Riemannian manifold $^{(3)}\mathcal{M}$, in which light geodesics are considered as spatial curve \cite{Gibbons:2008rj}. The metric $\gamma_{ij}$ and four-vector $N_{i}$ are given by
\begin{eqnarray}
dl^2=\gamma_{ij}dx^i dx^j&=&\frac{\Sigma^2}{\Delta(\Delta-a^2\sin^2\theta)}dr^2+\frac{\Sigma^2}{\Delta-a^2\sin^2\theta}d\theta^2\nonumber\\ 
&&+\left(r^2+a^2+\frac{2MG(r)ra^2\sin^2\theta}{\Delta-a^2\sin^2\theta}\right)\frac{\Sigma\sin^2\theta}{(\Delta-a^2\sin^2\theta)} d\phi^2,\nonumber\\
N_idx^i&=&-\frac{2MG(r)ar\sin^2\theta}{\Delta-a^2\sin^2\theta}d\phi,\label{metric3}
\end{eqnarray}
The Gauss-Bonnet theorem allows us to establish the relation between deflection angle of light and the surface integral of Gaussian curvature \cite{Ishihara:2016vdc, Ishihara:2016sfv, Gibbons:2008rj} via
\begin{equation}
\alpha_D=-\int\int_{{}_R^{\infty}\Box_{S}^{\infty}} K dS+\int_{S}^{R} k_g dl,\label{deflectionangle}
\end{equation}
where $K$ is the Gaussian curvature, $k_g$ is the geodetic curvature of light rays in $^{(3)}\mathcal{M}$, $dS$ is the area element of plane, and $dl$ is infinitesimal line element along light curve. To calculate the first integral in Eq.~(\ref{deflectionangle}), we consider the light propagation in the equatorial plane ($\theta=\pi/2$), therefore, the Gaussian curvature is simply defined as
\begin{eqnarray}
K&=&\frac{{}^{3}R_{r\phi r\phi}}{\gamma},\nonumber\\
&=&\frac{1}{\sqrt{\gamma}}\left(\frac{\partial}{\partial \phi}\left(\frac{\sqrt{\gamma}}{\gamma_{rr}}{}^{(3)}\Gamma^{\phi}_{rr}\right) - \frac{\partial}{\partial r}\left(\frac{\sqrt{\gamma}}{\gamma_{rr}}{}^{(3)}\Gamma^{\phi}_{r\phi}\right)\right),\label{GaussCurv}
\end{eqnarray}
where $\gamma=\det(\gamma_{ij})$. In the weak-field limit, upto the leading order terms, Eq.~(\ref{GaussCurv}) reads as
\begin{eqnarray}
K&=&\left(-\frac{2}{r^3}-\frac{6a^2}{r^5}+\frac{12\zeta}{r^5}+\frac{30\zeta a^2}{r^7} \right)M+\left(\frac{3}{r^4}-\frac{6a^2}{r^6}-\frac{22\zeta}{r^6} +\frac{20\zeta\gamma}{r^6}+\frac{15\zeta^2}{r^8} +\frac{36\zeta a^2}{r^8}+\frac{48\zeta a^2\gamma}{r^8}-\frac{54\zeta^2 a^2}{r^{10}}\right)M^2\nonumber\\
&&+ \left(\frac{8}{r^5}+\frac{12a^2}{r^7}-\frac{64\zeta}{r^7}-\frac{36\zeta\gamma}{r^7}-\frac{84a^2\zeta}{r^9}\right)M^3+\mathcal{O}\left(\frac{M^4}{r^6}, \frac{M^4a^2}{r^8},\frac{M^4\zeta}{r^{8}}\right).
\end{eqnarray}
Now, we calculate the integral of  Gaussian curvature over the closed quadrilateral as \cite{Ishihara:2016sfv}
\begin{equation}
\int\int_{{}_R^{\infty}\Box_{S}^{\infty}} K dS= \int_{\phi_S}^{\phi_R}\int_{\infty}^{r(\phi)} K \sqrt{\gamma}dr d\phi,\label{Gaussian}
\end{equation}
where $r(\phi)$ is the solution of the light orbit equation at the equatorial plane, using $r\equiv 1/u$ and $b=\xi$, Eqs.~(\ref{r}) and (\ref{phiuch}) yield 
\begin{equation}
\left(\frac{du}{d\phi}\right)^2=F(u),\label{orbit}
\end{equation}
with 
\begin{equation} F(u)=\frac{\left(1+a^2u^2-2M(1-\zeta u^2-M\zeta\gamma u^3)u\right)^2\Big(1+(a^2-b^2)u^2+2M(1-\zeta u^2-M\zeta\gamma u^3)u^3(a-b)^2\Big)}{[2M(1-\zeta u^2-M\zeta\gamma u^3)u(a-b)+b]^2}.\label{fu}
\end{equation}
In the weak-field approximation, we obtain the light orbit solution 
\begin{align}
u&=\frac{\sin\phi}{b} + \frac{M(1-\cos\phi)^2}{b^2}-\frac{2Ma(1-\cos\phi)}{b^3}-\frac{M^2(60\phi\,\cos\phi+3\sin3\phi-5\sin\phi)}{16b^3}\nonumber\\
&+\frac{a^2\sin^3\phi}{2b^3} +\mathcal{O}\left( \frac{M\zeta}{b^4},\frac{M\zeta\gamma}{b^4},\frac{M^3}{b^4}\right)
.\label{uorbit}
\end{align} 
For rotating ASG black hole spacetime, the surface integral of Gaussian curvature in Eq.~(\ref{Gaussian}) is calculated as
\begin{align}
\int\int K dS&=\left(\cos^{-1}bu_o+\cos^{-1}bu_s\right)\Big(\frac{15M^2}{4b^2}-\frac{4M^2a}{b^3}+\frac{105M^2a^2}{16b^4}-\frac{75M^2\zeta }{8b^4}- \frac{15M^2\zeta\gamma }{8b^4}+\frac{12M^2a\zeta}{b^5}\Big)\nonumber\\
&+ \left(\sqrt{1-b^2u_o^2}+\sqrt{1-b^2u_s^2}\right)\Big( \frac{2M}{b}+\frac{2Ma^2}{b^3}-\frac{8M\zeta}{3b^3}-\frac{32Ma^2\zeta}{5b^5}+\frac{77M^3}{6b^3}-\frac{6M^3a}{b^4}-\frac{248M^3\zeta\gamma}{15b^5}\Big) \nonumber\\
&+ \left(u_o\sqrt{1-b^2u_o^2}+u_s\sqrt{1-b^2u_s^2} \right)\Big( -\frac{M^2}{4b}+\frac{81M^2a^2}{16b^3}-\frac{51M^2\zeta}{8b^3}-\frac{15M^2\zeta\gamma}{8b^3}+\frac{12M^2a\zeta}{b^4}-\frac{95M^2a^2\zeta}{4b^5}-\frac{45M^2a^2\zeta\gamma}{8b^5}\Big)\nonumber\\
&+ \left(u_o^2\sqrt{1-b^2u_o^2} +u_s^2\sqrt{1-b^2u_s^2}  \right)\Big(\frac{Ma^2}{b}-\frac{4M\zeta}{3b}-\frac{16Ma^2\zeta}{5b^3}+\frac{84M^3a^2}{5b^3}-\frac{241M^3\zeta}{15b^3}+\frac{40M^3a\zeta\gamma}{3b^4}\Big)\nonumber\\
&+ \left(u_o^3\sqrt{1-b^2u_o^2} +u_s^3\sqrt{1-b^2u_s^2}  \right)\Big(\frac{15M^2a^2}{8b}-\frac{5M^2\zeta}{4b} -\frac{5M^2\zeta\gamma}{4b}+\frac{35M^2\zeta^2}{48b^3}-\frac{15M^2a^2\zeta\gamma}{4b^3}-\frac{211M^2a^2\zeta}{12b^3}\Big)\nonumber\\
&+\mathcal{O}\left(\frac{M^2\zeta^2}{b^6},\frac{M^4}{b^4} \right),\label{Gaussian1}
\end{align}
where $u_o$ and $u_s$ are the inverse of observer and source distances from the black hole and $\cos\phi_o=-\sqrt{1-b^2u_o^2},\; \cos\phi_s=\sqrt{1-b^2u_s^2}$. Furthermore, the geodesic curvature of light curve in $^{(3)}\mathcal{M}$ is defined as \cite{Ono:2017pie}
\begin{equation}
k_g=-\frac{1}{\sqrt{\gamma\gamma^{\theta\theta}}}N_{\phi,r}.
\end{equation}
For light propagating at the equatorial plane $(\theta=\pi/2)$, and considering the weak-field limit the geodesic curvature for metric (\ref{metric3}) of rotating ASG spacetime is calculated as
\begin{equation}
k_g=-\frac{2Ma}{r^3}-\frac{2M^2a}{r^4}+\frac{6M a \zeta}{r^5}+\frac{21M^3a}{r^5}+\frac{8M^2a\zeta}{r^6}+\frac{8M^2a\zeta\gamma}{r^6}-\frac{6M^2a\zeta^2}{r^8} +\mathcal{O}\left(\frac{M^3a\zeta}{r^7},\frac{M^3a\zeta\gamma}{r^7}\right).
\end{equation}
We must notice that the geodesic curvature identically vanishes for the non-rotating black hole spacetimes, thereby, the corresponding deflection in Eq.~(\ref{deflectionangle}) is solely given by the Gaussian curvature integral. The line element $dl$ along the light curve is given by
\begin{equation}
dl=\sqrt{\left( \gamma_{rr}\left(\frac{dr}{d\phi}\right)^2+\gamma_{\phi\phi}\right)}d\phi,
\end{equation}
this yields 
\begin{eqnarray}
\int_S^R k_g dl &=&
\left(\cos^{-1}bu_o+\cos^{-1}bu_s\right)\Big(-\frac{6M^2a}{b^3}-\frac{81M^2a\zeta }{4b^4}+\frac{3M^2a\zeta\gamma}{b^4}+\frac{6M^2a^2}{b^4}-\frac{9M^2a^2\zeta}{b^6}\Big)\nonumber\\ &+& \left(\sqrt{1-b^2u_o^2}+\sqrt{1-b^2u_s^2}\right)\Big(- \frac{2Ma}{b^2}+\frac{4Ma\zeta}{3b^4}+\frac{40M^3a^2}{b^5}-\frac{61M^3a}{2b^4}+\frac{69M^3a\zeta}{b^6}+\frac{112M^3a\zeta\gamma}{b^6}\Big)\nonumber\\
&+&\left(u_o\sqrt{1-b^2u_o^2}+u_s\sqrt{1-b^2u_s^2}\right)\Big(-\frac{2M^2a}{b^2}+\frac{2M^2a^2}{b^3}+\frac{27M^2a\zeta}{4b^4}+\frac{3M^2a\zeta\gamma}{b^3}-\frac{9M^2a^2\zeta}{b^5}\Big)\nonumber\\
&+& \left(u_o^2\sqrt{1-b^2u_o^2} +u_s^2\sqrt{1-b^2u_s^2}  \right)\Big(\frac{ 2Ma}{3b^2}-\frac{5M^3a}{b^2}+\frac{4M^3a^2}{b^3}+\frac{63M^3a\zeta}{2b^4}+\frac{56M^3a\zeta\gamma}{3b^3}-\frac{80M^3a^2\zeta\gamma}{3b^4}\Big)\nonumber\\
&+& \left(u_o^3\sqrt{1-b^2u_o^2} +u_s^3\sqrt{1-b^2u_s^2}  \right)\Big(\frac{2M^2a\zeta\gamma}{b}+\frac{3M^2 a\zeta }{b^2}-\frac{2M^2a^2\zeta}{b^3}\Big)+
\mathcal{O}\left(\frac{M^3a}{b^4},\frac{M^3a\zeta}{b^6}\right),
\label{geodesiccurvature}
\end{eqnarray}   
where we have considered the prograde motion of photon ($dl>0$), however, for retrograde motion ($dl<0$) we will get an extra -ve sign in Eq.~(\ref{geodesiccurvature}). 
Using Eqs.~(\ref{Gaussian1}) and (\ref{geodesiccurvature}) into Eq.~(\ref{deflectionangle}), we can compute the gravitational deflection angle of light in rotating ASG black hole spacetime for the finite-distance situations. However, in the very far distance limit, $u_o\to 0$ and $u_s\to 0$, the deflection angle, up to the leading order terms, yields
\begin{eqnarray}
\alpha_D&=&\left.\alpha_D\right|_{\text{Kerr}} -\left(\frac{16}{3b^3} -\frac{8a}{b^4}+ \frac{64 a^2}{5b^5} \right)\zeta M- \left(\frac{75\pi}{8b^4}+\frac{15\pi\gamma}{8b^4}-\frac{129\pi a}{4b^5}-\frac{3\pi a\gamma}{b^5}+\frac{53\pi a^2}{b^6}+\frac{45\pi a^2\gamma}{8b^6}\right)\zeta M^2\nonumber\\
&&-\left(\frac{144}{b^5}+\frac{192\gamma}{5b^5}\right)M^3\zeta+\mathcal{O}\left(\frac{Ma^3\zeta}{b^6},\frac{M^2\zeta^2}{b^6},\frac{M^3a\zeta}{b^6} \right),~\label{deflection}
\end{eqnarray}
this shows that the deflection angle is corrected by the ASG parameter $\zeta$. Here, $\left.\alpha_D\right|_{\text{Kerr}}$ is the Kerr deflection angle \cite{Ono:2017pie,Crisnejo:2019ril}
\begin{equation}
\left.\alpha_D\right|_{\text{Kerr}}=\left(\frac{4}{b} -\frac{4a}{b^2}+\frac{4a^2}{b^3}\right)M+\left( \frac{15\pi}{4b^2}-\frac{10\pi a}{b^3}\right)M^2 +\frac{128M^3}{3b^3} +\mathcal{O}\left(\frac{Ma^3}{b^4},\frac{M^2a^2}{b^4}, \frac{M^3a}{b^4}\right).
\end{equation}
The deflection angle of light for non-rotating ASG black hole is computed from Eq.~(\ref{deflection}) for $a=0$, which reads
\begin{equation}
\alpha_D=\left(\frac{4}{b} -\frac{16\zeta}{3b^3} \right)M+\left( \frac{15\pi}{4b^2}-\frac{75\pi\zeta}{8b^4}-\frac{15\pi\zeta\gamma}{8b^4}\right)M^2+ \left(\frac{128}{3b^3}-\frac{144\zeta}{b^5}-\frac{192\zeta\gamma}{5b^5}\right)M^3+  \mathcal{O}\left(\frac{M^4}{b^4} \right),
\end{equation}
and further retains the following value in the limiting case of $\zeta=0$ 
\begin{equation}
\alpha_D=\frac{4M}{b}+\frac{15\pi M^2}{4b^2}+\frac{128M^3}{b^3}  +\mathcal{O}\left(\frac{M^4}{b^4} \right),
\end{equation}
which gives the deflection angle for Schwarzschild black hole \cite{Virbhadra:1999nm}. 
\begin{figure}
\begin{tabular}{c c}
	\includegraphics[scale=0.75]{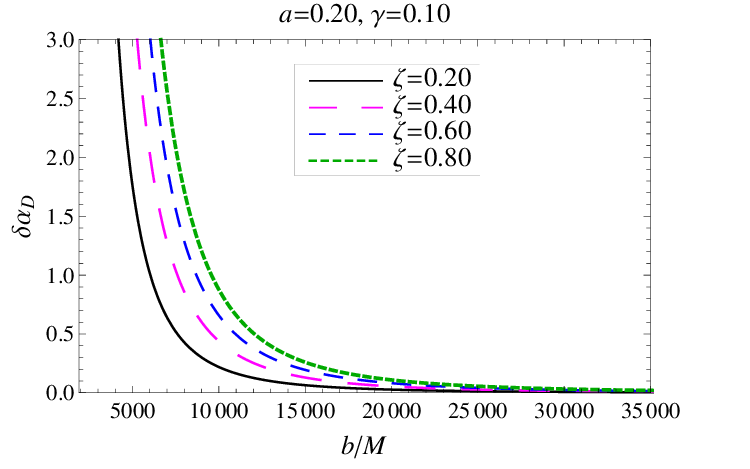}&
	\includegraphics[scale=0.75]{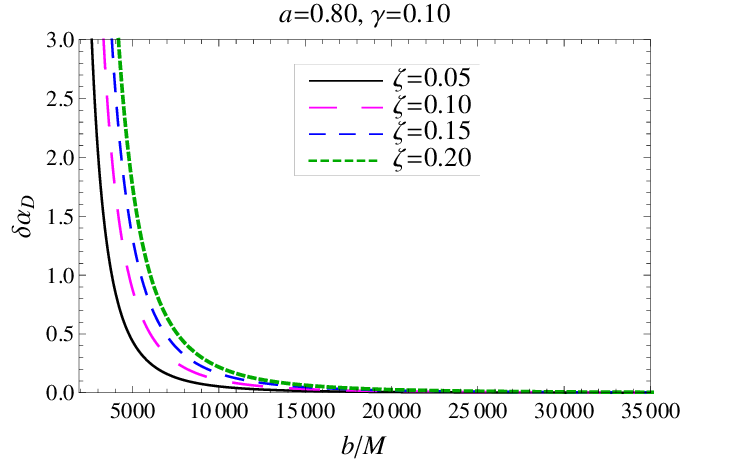}
\end{tabular}	
	\caption{The correction in the deflection angle $\delta \alpha_D=\left.\alpha_D\right|_{\text{Kerr}}-\alpha_D$ for different values of ASG parameters $\zeta$, and $a$; $\delta \alpha_D$ is in units of $\mu$as. }\label{lens2}
\end{figure}
It is evident that the deflection angle of light increases as the distance of closest approach to black hole decreases and eventually becomes unboundedly large at the photon sphere, which leads to the black hole shadow phenomenon \cite{Virbhadra:1999nm}. In Fig.~\ref{lens2}, we have plotted the correction to deflection angle due to ASG parameter $\zeta$ from the Kerr black hole value $\delta \alpha_D=\left.\alpha_D\right|_{\text{Kerr}}-\alpha_D$ in units of $\mu$as with varying $b$ and for various values of $\zeta$. It is evident that the effect of $\zeta$ on lensing angle subsided for sufficiently large impact parameter. Corrections in deflection angle from Kerr black hole $\delta \alpha_D=\left.\alpha_D\right|_{\text{Kerr}}-\alpha_D$ and from Schwarzschild black hole $\delta \alpha_D=\left.\alpha_D\right|_{\text{Schw}}-\alpha_D$ are summarized in Table \ref{T1} and Table \ref{T2}. This is evident that the correction increases with increasing $\zeta$. For given values of parameters, source, and observer's distances from the black hole, the rotating ASG black holes lead to smaller deflection angle than the Kerr ($\zeta=0$) and Schwarzschild ($\zeta=0,\, a=0$) black holes values, e.g., for $\zeta=0.20,\, a=0.40$ and $\gamma=0.10$, the corrections in deflection angle from Kerr and Schwarzschild black hole values are, respectively, $\delta\alpha_D=8.1306\,\mu$as and $\delta\alpha_D=36.6217\,$mas.

\begin{table}
	\begin{tabular}{ p{2cm} p{1.7cm} p{1.7cm} p{1.7cm} p{1.7cm} }
		\hline\hline
		$a $  & $\zeta=0.05$    & $\zeta=0.10$     & $\zeta=0.15$     &$ \zeta=0.20 $  \\  
		\hline\hline
		0.0&    2.03279& 4.06557& 6.09836& 8.13114 \\  \hline
		0.2&    2.03272& 4.06543& 6.09815& 8.13087 \\  \hline
		0.4&    2.03265& 4.0653& 6.09795& 8.1306   \\  \hline
		0.6&    2.03258& 4.06516& 6.09774& 8.13032 \\  \hline
		0.8&    2.03251& 4.06503& 6.09754& 8.13005 \\  \hline\hline
	\end{tabular}
	\caption{The corrections in the deflection angle $\delta\alpha_D =\left.\alpha_D\right|_{\text{Kerr}}-\alpha_D$ for Sgr A* with $b=3\times 10^3M$ and varying $\zeta$ and $a$; $\delta\alpha_D$ is in units of $\mu$as.}\label{T1}
\end{table}
\begin{table}
	\begin{tabular}{ p{2cm} p{1.7cm} p{1.7cm} p{1.7cm} p{1.7cm} }
		\hline\hline
		$a $  & $\zeta=0.05$    & $\zeta=0.10$     & $\zeta=0.15$     &$ \zeta=0.20 $  \\  
		\hline\hline
		0.0&    0.00203& 0.00406& 0.006096& 0.00813\\  \hline
		0.2&   18.3101& 18.3121& 18.3141& 18.3162 \\  \hline
		0.4&   36.6156& 36.6177& 36.6197& 36.6217 \\  \hline
		0.6&   54.9188& 54.9208& 54.9228& 54.9249 \\  \hline
		0.8&   73.2195& 73.2215& 73.2235& 73.2256 \\  \hline\hline
	\end{tabular}
	\caption{The corrections in the deflection angle $\delta\alpha_D =\left.\alpha_D\right|_{\text{Schw}}-\alpha_D$ for Sgr A* with $b=3\times 10^3M$ and varying $\zeta$ and $a$; $\delta\alpha_D$ is in units of $m$as.}\label{T2}
\end{table}
\section{Conclusion}\label{assect5}
The usual perturbative approach for quantization of gravity faces the problem of infinite numbers of diverging terms due to the dimensionality dependency of gravitational coupling constant. In the expedition of constructing a fully consistent theory of quantum gravity,  Weinberg \cite{S. Weinberg} made an important headway by fixing the running gravitational constant at UV scale. As far as the effects of asymptotic safety on black hole spacetimes are concerned, even in the infra-red limit, the black holes in ASG theory are significantly differ from their GR counterparts, though, the quantum effects diminish at asymptotically large distances. 

In this paper, we investigate whether these quantum corrections in the rotating ASG black holes propagate beyond the event horizon and left any observational imprints on the light propagation. In principle, light originating from the source get deflected when passing by a black hole, and the deflection angle depends upon the distance of closest approach. We consider the light propagation in the equatorial plane and using the Gauss-Bonnet theorem calculate the deflection angle in the weak-field limit. The geodesics curvature of light spatial curves in the optical geometry depends upon the ASG parameters and the black hole spin, however, it vanishes for non-rotating black hole. We analytically calculate the deflection angle by assuming that the observer and source are at finite distance from the black hole. For fixed values of black hole parameters, observer's and source's distances from the black hole the deflection angle of the light is modified by ASG parameters and generalize the Kerr deflection angle, however, the corrections in deflection angle are of order $\mathcal{O}(\mu$as), which is within the capabilities of current observational measurements. 
We also study the rotating ASG black hole shadow as perceived by a far distant observer sitting on the equatorial plane. The ASG parameters affect the near horizon geometry, and so does the black hole shadow. Shadow radius decreases with increasing $\zeta$, whereas the intrinsic distortion in rotating black hole shadows gradually increase with increasing $\zeta$. It is found that the effects of asymptotic safety on rotating black hole shadow is more prominent for higher values of the spin parameter, whereas for slowly rotating black holes it would be difficult to distinguish shadow in ASG theory from the corresponding one in standard GR. Hence, similar to the noncommutativity and the metric perturbations the quantum corrections manifested in the ASG theory also have observational effects in the shadow detection.

We constrain the ASG parameters with the aid of recent M87* black hole shadow observations. In turn, the M87* black hole shadow observables $(\Delta C, \theta_d)$ are found  to be consistent with the rotating ASG black holes shadows within the finite particular parameter space ($\zeta, a$), viz., the shadow angular diameter $\theta_d$ within the $1\sigma$ region leads to constraint on $\zeta$, viz., $\zeta\leq 0.1324$ for $\gamma=0.10$.

\section{Acknowledgments}
S.G.G. would like to thank DST INDO-SA bilateral project DST/INT/South Africa/P-06/2016, SERB-DST for the ASEAN project IMRC/AISTDF/CRD/2018/000042 and also IUCAA, Pune for the hospitality while this work was being done. R.K. would like to thank UGC for providing SRF.

\end{document}